\UseRawInputEncoding
\documentclass[final,5p,times,twocolumn,authoryear]{elsarticle}

\usepackage{amssymb}
\usepackage{lineno}
\usepackage{amsmath}
\usepackage{lipsum}
\usepackage{xcolor}
\usepackage[colorlinks]{hyperref}
\usepackage{siunitx}
\usepackage{graphicx}
\usepackage{titlesec}

\titleformat{\subsubsection}[runin]{\itshape}{\thesubsubsection}{3pt}{}[.]
\titlespacing*{\subsubsection}{\parindent}{1ex}{1em}

\newcommand{\change}[1]{{{#1}}}
\newcommand{\changetwo}[1]{{{#1}}}
\newcommand{\workflow}[1]{\texttt{{#1}}}

\newcommand{\DM}{\mathrm{DM}}

\newcommand{\f}{f}  % frequency symbol

\newcommand{\DMunits}{\si{pc\,\centi\meter^{-3}}}

\journal{Astronomy \& Computing}

\begin{document}

% \linenumbers

\begin{frontmatter}



\title{CELEBI: The CRAFT Effortless Localisation and Enhanced Burst Inspection Pipeline}


\author[icrar]{D.~R.~Scott\corref{cor1}}
\ead{danica.scott@postgrad.curtin.edu.au}
\cortext[cor1]{Corresponding author}

% alphabetical order for now
\author[harvard]{H.~Cho}
\author[mcgill]{C.~K.~Day}
\author[swin]{A.~T.~Deller}
\author[icrar]{M.~Glowacki}
\author[swin]{K.~Gourdji}
\author[atnf]{K.~W.~Bannister}
\author[icrar]{A.~Bera}
\author[atnf,astron,eric,apia]{S.~Bhandari} 
\author[icrar]{C.~W.~James}
\author[swin]{R.~M.~Shannon}

\affiliation[icrar]{organization={International Centre for Radio Astronomy Research},
            addressline={Curtin University}, 
            city={Bentley},
            postcode={6102}, 
            state={WA},
            country={Australia}}

\affiliation[harvard]{organization={Center for Astrophysics | Harvard \& Smithsonian},
            addressline={60 Garden Street},
            city={Cambridge},
            postcode={02138},
            state={MA},
            country={United States of America}}

\affiliation[mcgill]{organization={Department of Physics},
            addressline={McGill University},
            city={Montreal},
            postcode={H3A 2T8},
            state={Quebec},
            country={Canada}}

\affiliation[swin]{organization={Centre for Astrophysics and Supercomputing},
            addressline={Swinburne University of Technology},
            city={Hawthorn},
            postcode={3122},
            state={VIC},
            country={Australia}}

\affiliation[atnf]{organization={CSIRO Space and Astronomy, Australia Telescope National Facility},
            addressline={PO Box 76},
            city={Epping},
            postcode={1710},
            state={NSW},
            country={Australia}}
            
\affiliation[astron]{organization={ASTRON, Netherlands Institute for Radio Astronomy},
            addressline={Oude Hoogeveensedijk 4},
            city={Dwingeloo},
            postcode={7991 PD},
            country={The Netherlands}}
            
\affiliation[eric]{organization={Joint institute for VLBI ERIC},
            addressline={Oude Hoogeveensedijk 4},
            city={Dwingeloo},
            postcode={7991 PD},
            country={The Netherlands}}
            
\affiliation[apia]{organization={Anton Pannekoek Institute for Astronomy, University of Amsterdam},
            addressline={Science Park 904},
            city={Amsterdam},
            postcode={1098 XH},
            country={The Netherlands}}

% \affiliation[]{organization={},

% \author{others}

% \author[inst2]{Author Two}
% \author[inst1,inst2]{Author Three}

% \affiliation[inst2]{organization={Department Two},%Department and Organization
%             addressline={Address Two}, 
%             city={City Two},
%             postcode={22222}, 
%             state={State Two},
%             country={Country Two}}


% \begin{highlights}
% \end{highlights}
\begin{abstract}
Fast radio bursts (FRBs) are being detected with increasing regularity. However, their spontaneous and often once-off nature makes high-precision burst position and frequency-time structure measurements difficult without specialised real-time detection techniques and instrumentation. The Australian Square Kilometre Array Pathfinder (ASKAP) has been enabled by the Commensal Real-time ASKAP Fast Transients Collaboration (CRAFT) to detect FRBs in real-time and save raw antenna voltages containing FRB detections. We present the CRAFT Effortless Localisation and Enhanced Burst Inspection pipeline (CELEBI), an automated \change{offline} software pipeline that extends CRAFT's existing software to process ASKAP voltages in order to produce sub-arcsecond precision localisations and polarimetric data at time resolutions as fine as 3\,ns of FRB events. We use Nextflow to link together Bash and Python code that performs software correlation, interferometric imaging, and beamforming, making use of common astronomical software packages.
\end{abstract}
\begin{keyword}
Fast Radio Bursts \sep Radio Interferometry \sep Astronomy Software 

\end{keyword}

\end{frontmatter}

\section{Introduction}
\label{sec:intro}

Fast radio bursts (FRBs) are micro- to millisecond duration radio transients \citep{lorimer_bright_2007, thornton_population_2013}. Known to be extragalactic, they are extremely energetic \citep{bhandari_host_2020}. Although a Galactic magnetar is known to have produced FRB-like emission \citep{chimefrb_collaboration_bright_2020, bochenek_fast_2020}, no general emission mechanism nor progenitor has been identified, and it is possible that more than one progenitor type contributes to the observed population. Only a small fraction of FRB sources have been observed to repeat (\citet{chimefrb_collaboration_chimefrb_2023} report a repeater fraction tending to $2.6^{+2.9}_{-2.6}\%$) and there are indications of intrinsic differences between repeating and non-repeating FRBs \citep{pleunis_fast_2021}.

In order to gain greater insight into the nature of FRBs, their emission mechanisms, progenitors, and host environments, and to use them as probes of cosmological parameters \citep{james_measurement_2022} and extragalactic matter distributions \citep{macquart_census_2020} it is highly desirable and often necessary to identify their host galaxies and measure the polarimetric morphologies of the bursts themselves at high temporal and spectral resolutions. For example, the current sample of host galaxies does not yet point to any preferred progenitor class \citep{bhandari_characterizing_2022} but patterns may emerge as the sample grows, and high-time resolution measurements can constrain the size of the emission region and therefore emission mechanisms \citep{nimmo_burst_2021}.

\changetwo{
The computational load associated with FRB searches scales with increasing spatial, time, and frequency resolutions, so compromises must be made to make the load manageable. In many cases, these compromises result in an inability to localise most FRB sources with sufficient precision to identify a host galaxy, measure bursts with sufficient temporal or spectral resolutions to make detailed inferences on the emission mechanism, or measure the polarimetric properties of bursts.

Making associations between FRBs and their host galaxies, whilst still restricted to a small fraction of the detected FRB population, has been accomplished by a number of different radio interferometers. For relatively nearby FRBs, lower localisation precision is needed given the lower sky density of potential host galaxies, and a number of repeating FRBs detected by the Canadian Hydrogen Intensity Mapping Experiment (CHIME) have been associated on the basis of moderate ($\lesssim$arcminute level) localisation precision afforded by detailed beam modelling \citep{bhardwaj_nearby_2021, bhardwaj_local_2021, michilli_sub-arcminute_2022, ibik_proposed_2023}. Other interferometers operating at higher frequency and with longer baselines have successfully associated non-repeating FRBs to their host galaxies using data products formed in real time. The VLA has localised one apparently non-repeating FRB to a host galaxy \citep{law_distant_2020} using \textit{realfast}, a real-time imaging FRB search pipeline \citep{law_realfast_2018}. MeerKAT has localised two apparently non-repeating FRBs, detected in an incoherent beam of $\sim1.3\,\si{\deg}^2$, to sub-arcsecond precision in images made from standard correlation products and subsequently identified their host galaxies \citep{caleb_discovery_2023, driessen_frb_2023}, and a method for localising transients detected across multiple MeerKAT tied-array beams is described by \cite{bezuidenhout_tied-array_2023}. The Westerbork Synthesis Radio Telescope's Apertif system is able to detect FRBs in real time and consequently localise, but without the spatial resolution necessary to identify a host galaxy \citep{connor_bright_2020}.

Another method for localising non-repeating FRBs involves a real-time search that triggers the capture of high-resolution data for offline processing that otherwise would not have been saved due to their large volume. The Deep Synoptic Array is able to capture FRBs in voltages and interferometrically localise them to arcsecond precision, allowing identification of host galaxies \citep{ravi_fast_2019, ravi_deep_2022, ravi_deep_2023}. While it is yet to localise a non-repeating FRB to a host galaxy, CHIME has a pipeline for real-time FRB detection and capture of baseband data, permitting sub-arcminute FRB localisations and measurement of burst profiles at microsecond time resolution from single detections \citep{michilli_analysis_2021}.
}

The Australian Square Kilometre Array Pathfinder (ASKAP, \citealt{hotan_australian_2021}) has been enabled by the Commensal Real-time ASKAP Fast Transients Collaboration (CRAFT) to detect FRBs in real-time and save raw antenna voltages of FRB detections. This permits sub-arcsecond-precision localisation of FRBs, including non-repeating FRBs, via interferometric imaging, precise enough to identify a host galaxy and often a position within that galaxy \citep{bannister_single_2019, prochaska_low_2019, macquart_census_2020, heintz_host_2020, fong_chronicling_2021, bhandari_non-repeating_2022, bhandari_characterizing_2022, ryder_probing_2022}, and polarimetric measurements at time resolutions as fine as 3\,\si{\nano\second} \citep{cho_spectropolarimetric_2020, day_high_2020}. However, to date, the post-processing of triggered FRB data products has been handled by an ensemble of processing scripts that are manually sequenced and which require significant human quality control.  This process is time consuming and potentially error-prone, making it unsuitable for future FRB surveys with ASKAP with higher detection rates (as envisaged for the forthcoming ``CRACO'' coherent detection system; Bannister et al., in prep.).

This paper describes the CRAFT Effortless Localisation and Enhanced Burst Inspection pipeline (CELEBI)\change{, which has been made publicly available on Github\footnote{\href{https://github.com/askap-craco/CELEBI}{github.com/askap-craco/CELEBI/}}}. CELEBI is an automated \change{offline} software pipeline that extends existing CRAFT post-processing code \citep{bannister_single_2019, cho_spectropolarimetric_2020} with new functionality and improved monitoring and control to produce sub-arcsecond-precision localisations and high-time resolution data products of FRBs detected with ASKAP \change{with minimal human oversight and direct intervention}. \S\ref{sec:overview} gives a high-level overview of the pipeline's structure and algorithm. \S\ref{sec:localisation} describes in detail the processes performed by CELEBI to produce FRB localisations, each subsection corresponding directly to one of CELEBI's processes. \S\ref{sec:htr} similarly describes the processes that produce high-time resolution polarimetric data for FRB detections. \S\ref{sec:discussion} gives a summary and discusses the improvements to CRAFT's voltage processing produced by CELEBI, as well as future improvements to the pipeline. \change{We use FRB190711 \citep{day_high_2020} as an example case throughout to demonstrate consistency with previously-published results.}
\section{Overview}
\label{sec:overview}

\subsection{Input data format}
\label{sec:data_format}
CELEBI's primary inputs are sets of voltages acquired by the simultaneous freezing and downloading (``dumping'') of the contents of a 3.1\,s-duration ring buffer for each ASKAP antenna. The buffers record complex-valued electric field samples across a 336-MHz bandwidth in both orthogonal linear polarisations of each beam of an antenna's phased-array feed (PAF), although only the data for the beam in which the desired target is detected in is saved. Upon the real-time detection of an FRB, the voltages are dumped with sufficiently low latency to capture the FRB \citep{bannister_single_2019}. Voltages are then obtained for two other sources: a ``flux" calibrator (a bright continuum source: typically PKS~0408$-$65 or PKS~1934$-$63), and a polarisation calibrator (a bright, highly linearly polarised pulsar: typically Vela or PSR~J1644$-$4559). As described below, these two datasets are employed to derive the necessary calibration terms that enable astrometrically and polarimetrically correct images and time series to be formed for the FRB. \change{The data for the calibrators are typically significantly separated from the FRB, both temporally and spatially. This will produce systematic errors in the images produced from the FRB voltages as the calibration solutions are applied. We account for this using the method described by \cite{day_astrometric_2021} in process \ref{proc:find_offset}. With regards to measuring the polarimetric properties of the polarisation calibrator, ASKAP is sufficiently stable that we do not expect the polarisation calibration solutions derived to be invalid when applied to the FRB data.}

\begin{figure}
    \centering
    \includegraphics[scale=1]{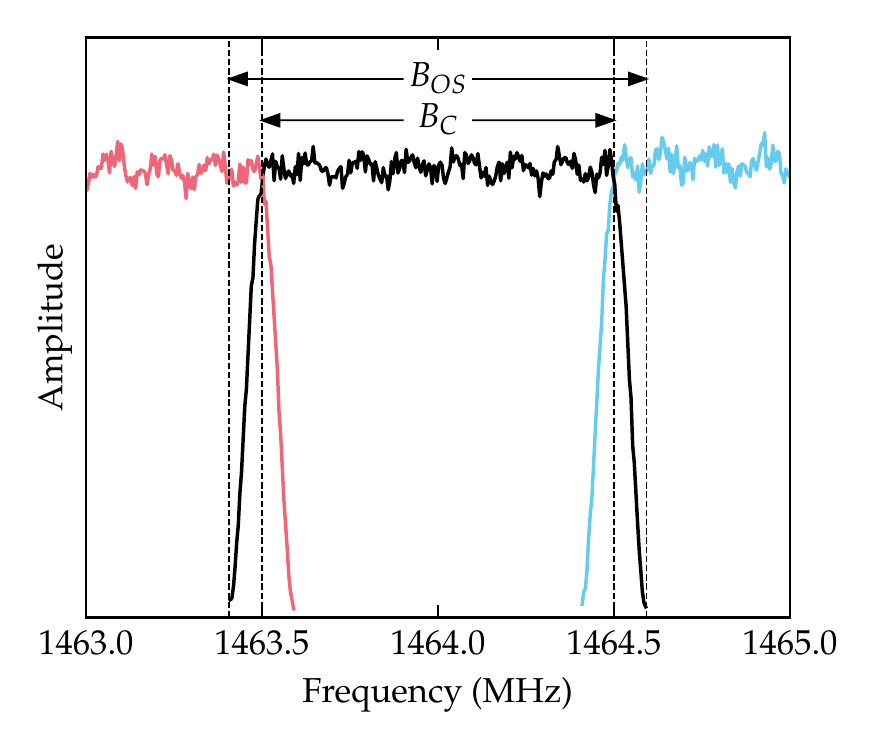}
    \caption{Amplitude response as a function of frequency within three adjacent coarse channels pre-PFB inversion, averaged over 4096 fine channels. $B_C=1\,\si{\mega\hertz}$ is the coarse channel bandwidth and $B_{OS}=(32/27)B_C\approx1.19B_C$ is the oversampled bandwidth. A predictable ripple in the passband amplitude is present due to the response function of the PFB used to form these 1\,MHz channels.}
    \label{fig:oversamp_chan}
\end{figure}

ASKAP's polyphase filterbanks (PFBs) produce oversampled ``coarse'' channels, meaning adjacent channels overlap slightly. Each coarse channel is composed of many ``fine'' channels, the precise number of which is dependent on the amount of data read from file which is dynamically determined during processing. Each PFB produces data across 784 coarse channels, each separated by $B_C=1\,\si{\mega\hertz}$. The oversampled bandwidth of each channel is $B_{OS} = (32/27)B_C\approx1.19B_C$. Each channel has a region of locally rippled but overall constant frequency response with width $B_C$, and an oversampled region of width $(B_{OS}-B_C)/2$ on either side that tapers off, as seen in Figure \ref{fig:oversamp_chan}, which shows the fine spectrum amplitudes in three adjacent coarse channels. Of these \change{784 coarse channels}, 336 are available for real-time analysis by the incoherent sum (ICS) pipeline \citep{bannister_single_2019} and recorded to voltage buffers for offline analysis.

Voltages are stored as 8-bit complex numbers (4 bit real, 4 bit imaginary) in ``VCRAFT" files, each of which contains two sets of four 1\,MHz oversampled coarse channels. Each set of four channels are internally contiguous in frequency, but the two sets within a VCRAFT file are not necessarily contiguous with each other. \change{Each VCRAFT file was produced by one of six FPGAs on one of seven processing cards for a total of 42 VCRAFT files per polarisation per antenna, and a total bandwidth of 336\,MHz. The headers for these VCRAFT files contain frequency and timestamp information. The observing band is typically one of the ``lower'' (central frequency 863.5\,MHz), ``mid'' (central frequency 1271.5\,MHz), or ``upper'' (central frequency 1632.5\,MHz) bands, with a fixed bandwidth of 336\,MHz. We make use of the CRAFT utilities\footnote{\href{https://github.com/askap-craco/craft}{github.com/askap-craco/craft}} to read and manipulate VCRAFT data.}

As well as the raw voltage data, a set of metadata associated with the FRB trigger is also provided as input. This includes the real-time search candidate that triggered the voltage dump, the parameters of the observation the detection was made in, and a preliminary FRB position derived from multibeam analysis with a precision of a few arcminutes.

The candidate that triggered the voltage dump is stored as a text file containing the candidate's signal-to-noise ratio (S/N), arrival time, dispersion measure (DM), and boxcar width. This candidate is the first that passes the S/N, DM, and width filters of the real-time search and as such the parameters are considered preliminary.

The observational parameters provided include the name, location, and fixed delay associated with each ASKAP antenna.

\subsection{Algorithm overview}
CELEBI's primary functionalities are high-precision FRB localisation via interferometric imaging and obtaining high-time resolution data products via tied-array beamforming. It is constructed from Python and Bash scripts linked into a single Nextflow\footnote{\href{https://nextflow.io}{nextflow.io}} pipeline \citep{di_tommaso_nextflow_2017}. Nextflow manages data flow between processes, automation of process execution \& parallelisation, and submission of processes to supercomputing resources. CELEBI makes use of the Astronomical Image Processing System (AIPS, \citealt{greisen_aips_2003}), ParselTongue\footnote{\href{https://www.jive.eu/jivewiki/doku.php?id=parseltongue:parseltongue}{jive.eu/jivewiki/doku.php?id=parseltongue:parseltongue}}, the Common Astronomy Software Applications (CASA, \citealt{casa_team_casa_2022}), the CASA Analysis Utilities\footnote{\href{https://casaguides.nrao.edu/index.php/Analysis_Utilities}{casaguides.nrao.edu/index.php/Analysis\_Utilities}}, CRAFT utilities, DiFX \citep{deller_difx-2_2011} \change{and psrvlbireduce\footnote{\href{https://github.com/dingswin/psrvlbireduce}{github.com/dingswin/psrvlbireduce}}}, as well as the Python libraries Numpy \citep{harris_array_2020}, Scipy \citep{virtanen_scipy_2020}, Matplotlib \citep{hunter_matplotlib_2007}, Astropy \citep{robitaille_astropy_2013, astropy_collaboration_astropy_2018, astropy_collaboration_astropy_2022}, and Astroquery \citep{ginsburg_astroquery_2019}. \change{These dependencies, as well as example execution commands for CELEBI, are listed in the README of CELEBI's github repository.}

\begin{figure}
    \centering
    \includegraphics[scale=0.5]{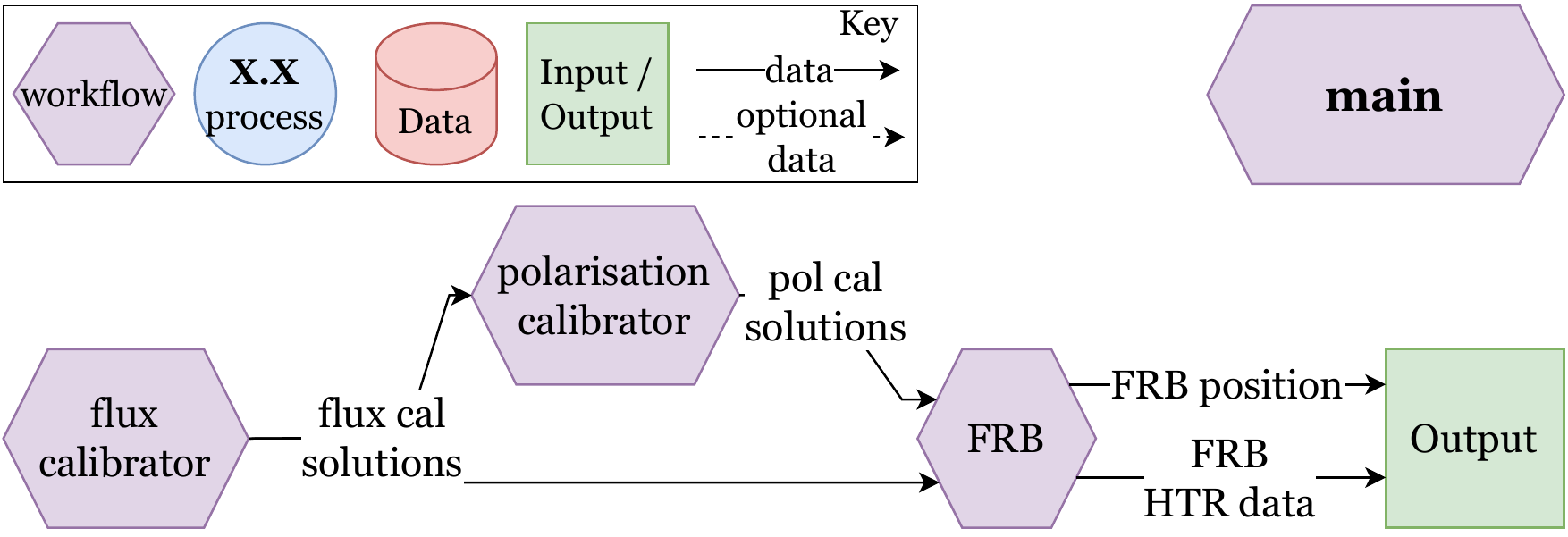}
    \caption{Top-level pipeline DFD. Inset: key for DFD symbols.}
    \label{fig:main}
\end{figure}

CELEBI is split into three major workflows, each associated with a set of voltages: \workflow{fluxcal}, \workflow{polcal}, and \workflow{FRB}. These \change{are} linked by a \workflow{main} workflow, a data-flow diagram (DFD) of which is shown in Figure \ref{fig:main}.

\begin{figure*}
    \centering
    \includegraphics[scale=0.5]{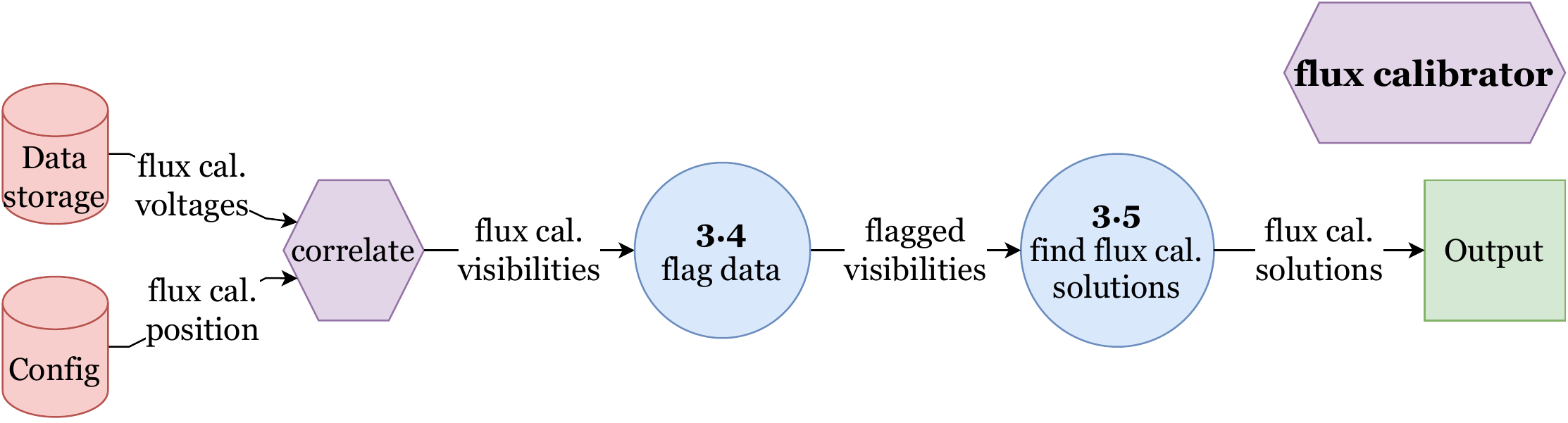}
    \caption{\workflow{fluxcal} workflow DFD.}
    \label{fig:fluxcal}
\end{figure*}

\workflow{fluxcal} (Figure \ref{fig:fluxcal}) produces frequency-dependent complex gain solutions \change{and flux scaling} from the ``fluxcal"  calibrator voltages. These solutions are used in \workflow{polcal} and \workflow{FRB} for imaging and beamforming. After the application of these solutions, each polarisation of each antenna should be (independently) correctly calibrated, meaning that the different antennas are correctly aligned and the amplitude scale is correctly placed in units of Janskys. However, the two polarisations may still be offset in delay or phase with respect to each other.

\begin{figure*}
    \centering
    \includegraphics[scale=0.5]{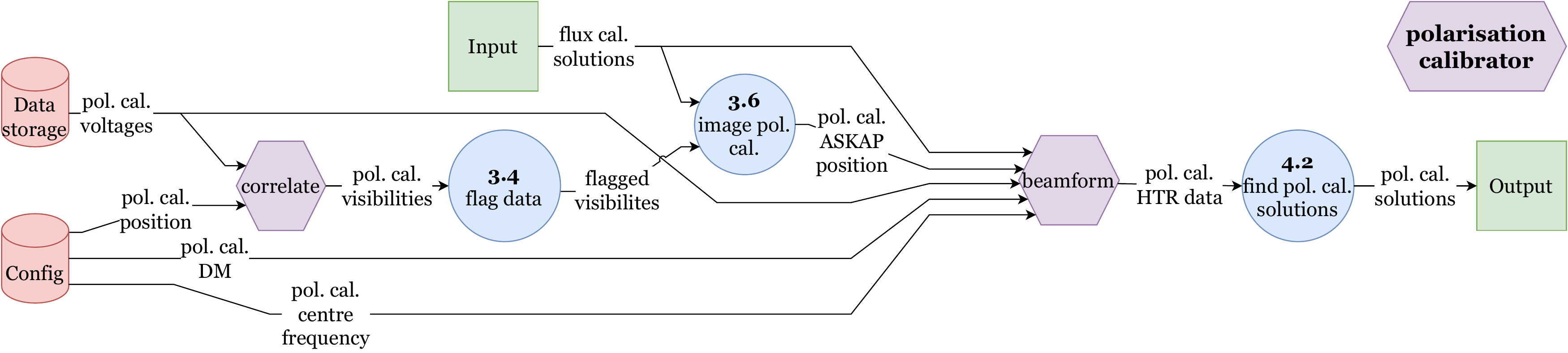}
    \caption{\workflow{polcal} workflow DFD.}
    \label{fig:polcal}
\end{figure*}

\workflow{polcal} (Figure \ref{fig:polcal}) produces polarisation calibration solutions that correct for the small delay and leakage {\em between} the two nominally orthogonal polarisation bases of the ASKAP antennas from the polarisation calibrator voltages.

\begin{figure*}
    \centering
    \includegraphics[scale=0.5]{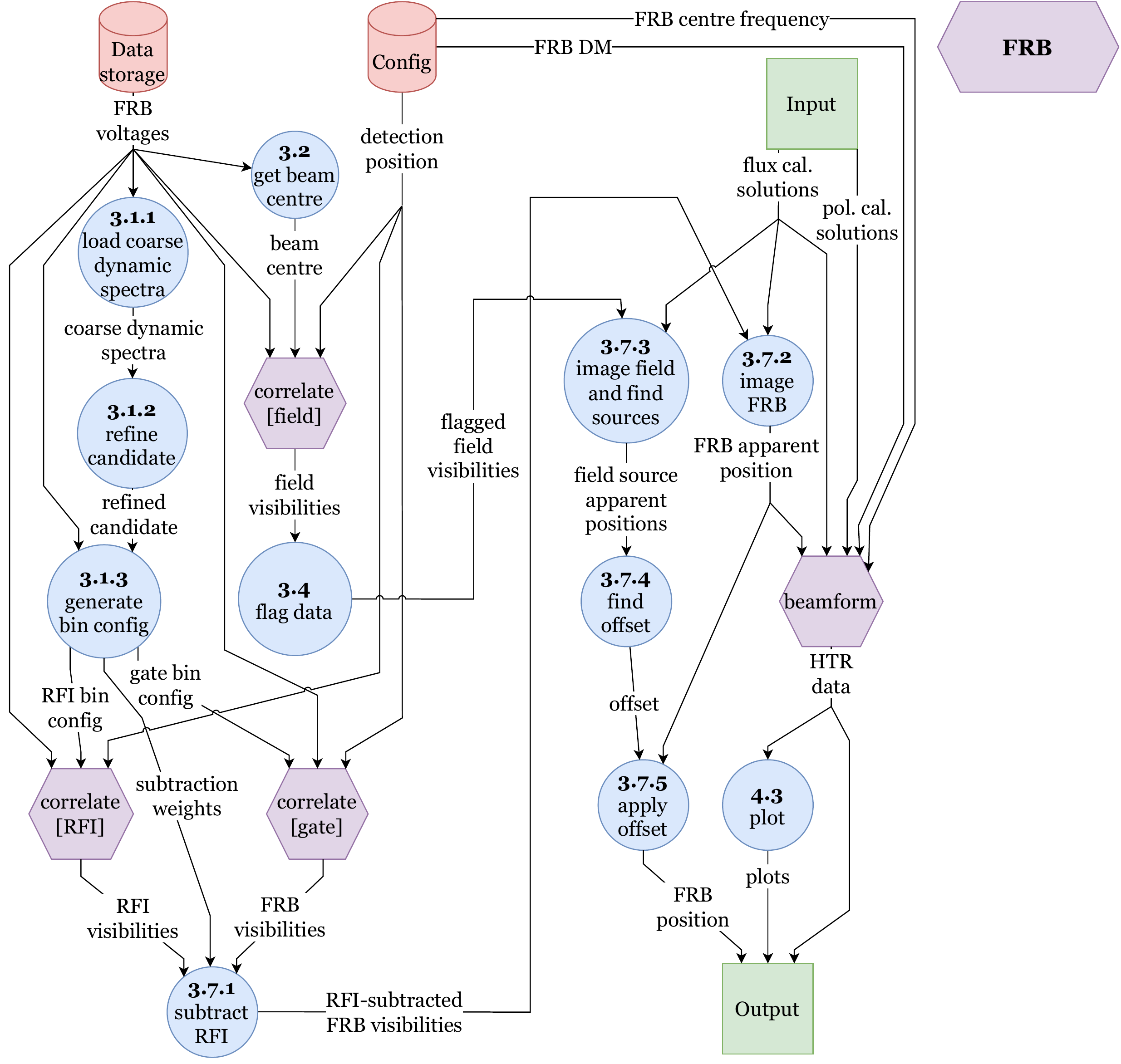}
    \caption{\workflow{FRB} workflow DFD.}
    \label{fig:frb}
\end{figure*}

\workflow{FRB} (Figure \ref{fig:frb}) takes the solutions from the other major workflows and processes the FRB voltages to determine a sub-arcsecond-precision localisation of the FRB and produce a set of high-time resolution polarimetric data.

\section{FRB and polarisation calibrator localisation}
\label{sec:localisation}

In each of the major workflows, the voltages are correlated in software using the \workflow{correlate} workflow (\S\ref{wf:correlate}, Figure \ref{fig:correlation}) to produce visibilities, which are then processed in specific ways depending on the purpose of each data set. 

\workflow{fluxcal} uses its visibilities to derive frequency dependent complex gain solutions (process \ref{proc:find_flux_soln}), which are provided to the other major workflows, where they are used first in imaging, then passed as input to the \workflow{beamform} workflow (\S\ref{wf:beamform}, Figure \ref{fig:beamform}).

\workflow{polcal} images its visibilities to localise the calibrator (process \ref{proc:image_polcal}) to provide its apparent position to \workflow{beamform}, which ultimately produces high-time resolution polarimetric data which is referenced to a model in order to derive polarisation calibration solutions (process \ref{sec:polcal}). These solutions are then passed on to \workflow{FRB}.

\workflow{FRB} performs three different modes of correlation: field, gate, and RFI (radio-frequency interference). The field mode is a simple correlation of the full 3.1\,\si{\second} voltage dump, used to image the field around the FRB position (process \ref{proc:image_field}). This is used to derive an astrometric correction for a systematic offset in apparent sky position introduced by the difference in observation time and direction between the FRB detection and later flux calibrator observation (process \ref{proc:find_offset}) \citep{day_astrometric_2021}. The gate mode selects a small, frequency-dependent time window of data for correlation that matches the arrival time of the FRB itself in order to create an image of only the FRB and maximising signal-to-noise. The RFI mode similarly restricts the data used, but instead includes only small (frequency-dependent) time windows immediately before and after the FRB. The RFI visibilities are subtracted from the FRB visibilities in order to eliminate RFI without removing entire channels that may contain important components of the FRB signal (process \ref{proc:subtract_rfi}). On these millisecond timescales, RFI contributions to visibilities remain approximately constant, meaning that subtracting an appropriately scaled (in amplitude) copy of the RFI dataset from the FRB dataset can effectively mitigate RFI in frequency ranges that would otherwise be flagged and lost. The FRB is imaged, its position fit (process \ref{proc:image_FRB}), and the offset derived from the field image applied to calculate the final astrometrically-corrected FRB position (process \ref{proc:apply_offset}).

\subsection{Incoherent search}

CRAFT's real-time FRB search \change{\citep{bannister_single_2019}} tests a grid of DM and arrival time values, meaning that a single FRB event typically produces several candidates with a S/N above the threshold for triggering a voltage dump. Because the dump is triggered on the first candidate seen above threshold, the measured DM and arrival time of this candidate is likely not to be the combination that produces the highest S/N. In order to optimise S/N when imaging the FRB, we first refine the FRB candidate to measure a more accurate DM and arrival time by performing an incoherent search on the voltages following the same principles as CRAFT's real-time detection system.

\subsubsection{Load coarse dynamic spectra}
\label{proc:coarse_dynspecs}
This process loads the voltages from the FRB data set and constructs a ``coarse'' dynamic spectrum. \change{The frequency resolution is chosen to be 1\,MHz for simplicity of implementation (the data is already channelised to 1 MHz), and the time resolution is chosen to be 1\,ms as this is fine enough to allow sufficient precision in refining the FRB candidate, while keeping the processing time of candidate refinement low.} 

\change{To construct a coarse dynamic spectrum, we load the coarse-channelised voltages and for each coarse channel:
\begin{enumerate}
    \item Fast Fourier Transform (FFT) into the spectral domain.
    \item Trim the oversampled regions (Figure \ref{fig:oversamp_chan}).
    \item Inverse FFT to obtain a 1\,\si{\micro\second} time resolution complex voltage time series.
    \item Take the square of the absolute power of this time series to obtain a power time series.
    \item Reduce the time resolution (``time scrunch'') to 1\,\si{\milli\second} by summing blocks of 1000 time samples.
\end{enumerate}
The time scrunched power time series in each coarse channel is then arranged into a two-dimensional dynamic spectrum, which is passed as output of the process.}

We account for geometric delays in signal arrival time between antennas by applying offsets in time in the data \change{as it is} read from the VCRAFT files based on an interferometer model calculated using the DiFX program \texttt{difxcalc} \citep{2016ivs..conf..187G} for the initial rough FRB position.

An instance of this process is run for each unique polarisation-antenna pair. 

This process also calculates \change{and outputs} the time axis of the dynamic spectra in units of Modified Julian Day (MJD) based on the start time of the data in the VCRAFT headers and the geometric delays.

\subsubsection{Refine candidate}
\label{proc:refine_cand}
This process sums the power dynamic spectra across both polarisations and all antennas as output by \ref{proc:coarse_dynspecs}, then searches the resulting ICS dynamic spectrum for a single dispersed pulse. We search over DMs in a configurable range of DM values, defaulting to $\pm10$\,\DMunits\ around the detection candidate's DM with a step size of 0.01\,\DMunits. For each DM, we incoherently dedisperse the ICS dynamic spectrum by shifting each coarse channel by an integer number of time samples. The number of 1\,\si{\milli\second} samples a coarse channel of central frequency $\f$ is shifted in the direction of increasing time is given by

\begin{equation}
   t_\mathrm{shift}(\DM, \f)
        = \left\lfloor k_\DM \DM \left(\f_0^{-2} - \f^{-2}\right)\cdot1000 \right\rfloor,
\end{equation}

\noindent where $k_\DM = (2.41\times10^{-4})^{-1}\,\mathrm{MHz}\,\mathrm{cm}^3\,\mathrm{pc}^{-1}$, $\f_0$ is a reference frequency, which we choose as the central frequency of the lowest-frequency coarse channel, \change{and the factor of 1000 converts from seconds to milliseconds}. Note that this choice of reference frequency results in the lowest-frequency coarse channel not being shifted at all, and as such we can use the MJD time array produced by process \ref{proc:coarse_dynspecs} to measure the arrival time of the burst at the bottom of the observing band. We then sum this dedispersed dynamic spectrum along the frequency axis to get a 1\,\si{\milli\second}-resolution dedispersed profile.

To improve the S/N of the FRB in the event that it is spread out over multiple time samples, either due to the burst's intrinsic width or dispersive smearing within channels, we smooth the profile by convolving with top-hat functions with widths between 1 and 10 samples. We then calculate the S/N of each sample in each smoothed profile by dividing each profile by its standard deviation. We take the DM, time, and width \change{corresponding to} the maximum S/N value \change{($\mathrm{DM_{opt}}$, $t_\mathrm{opt}$, and $w_\mathrm{opt}$ respectively)} and create a refined candidate file with these values updated.

\subsubsection{Generate bin configs}
\label{proc:generate_binconfig}

The gate and RFI correlation modes performed in \workflow{FRB} require the specification of time and frequency dependent weights that are used to select only certain windows of data for correlation and imaging. The windows are shaped according to the expected time delay due to dispersion across the band based on the arrival time and DM of the refined candidate generated by process \ref{proc:refine_cand}. These take the form of ``bin config'' files, that record the windows and their weights, and a ``polyco'' file that records a reference time \& frequency and the DM of the FRB.

\begin{figure}
    \centering
    \includegraphics[width=\columnwidth]{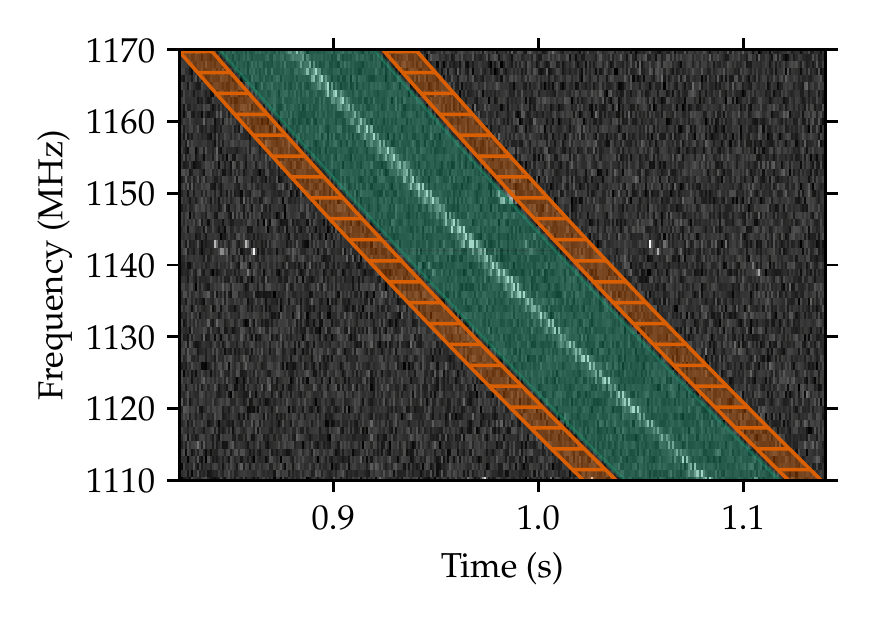}
    \caption{\change{Cropped FRB190711} dynamic spectrum (time scrunched to 1\,\si{\milli\second}) with example gate bin region (green) and RFI bins (orange hatched) overlaid. The green region is split into seven equal-width bins.}
    \label{fig:bins_dynspec}
\end{figure}

Figure \ref{fig:bins_dynspec} illustrates how the bins are defined for the gate and RFI modes. \change{The gate bins are defined by taking a region of width $w_\mathrm{opt}+70\,\si{\milli\second}$, with a dispersive sweep corresponding to $\mathrm{DM_\mathrm{opt}}$, and centred on $t_\mathrm{opt}$ at the bottom of the band. This region is divided into seven bins.} The central bin is expected to be the only bin to contain the FRB, but all are imaged in case part or all of the FRB signal falls outside of the central bin due to small unexpected errors in identifying the burst arrival time.

The RFI bins are each 16\,\si{\milli\second} wide and follow the same dispersive sweep as the \change{gate region}, leaving a 4\,\si{\milli\second} buffer on either side of the gate \change{region}. This width was chosen as a compromise between the counter-posed goals of minimising the noise contribution (which favours a longer duration) and measuring the RFI environment as close in time as possible to the FRB itself (which favours a shorter duration).

\subsection{Get beam centre}
We wish to centre the field image on the beam centre, rather than the preliminary FRB position. This process parses the FRB voltage headers in order to obtain the sky coordinates of the beam centre, which are passed to the field mode's \workflow{correlate} instance.

\subsection{Correlate}
\label{wf:correlate}

\begin{figure}
    \centering
    \includegraphics[scale=0.5]{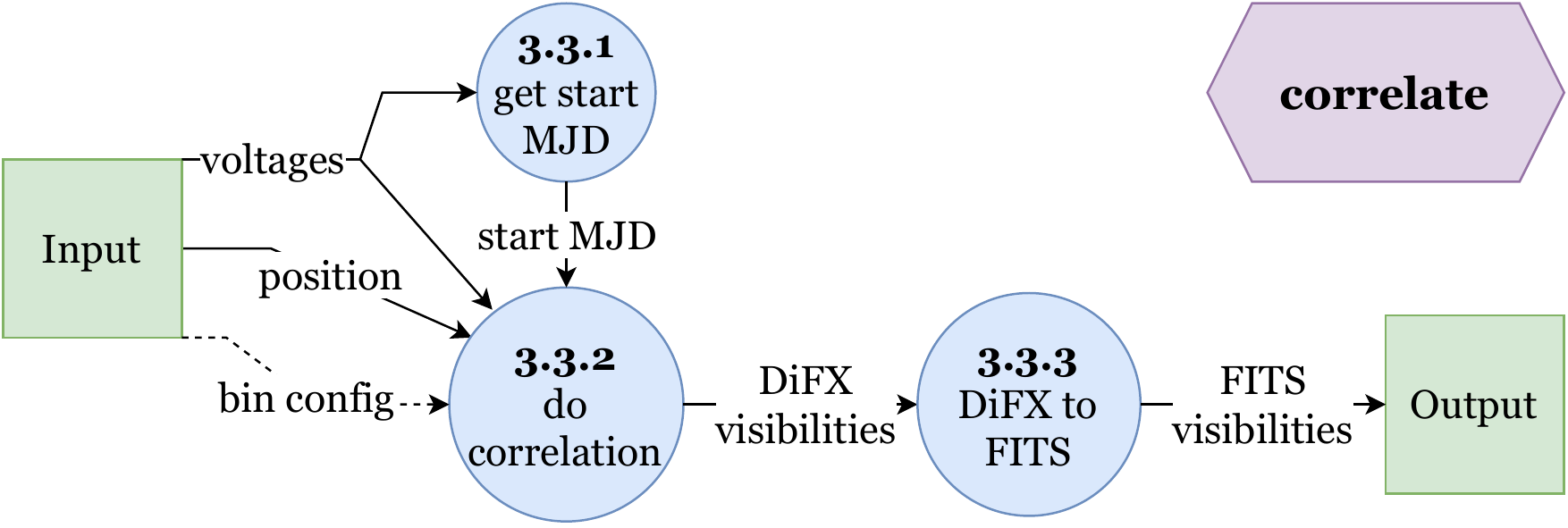}
    \caption{Correlation workflow DFD.}
    \label{fig:correlation}
\end{figure}

The \texttt{correlate} workflow takes as input a VCRAFT voltage dataset (as described in \S\ref{sec:data_format}) and an optional bin configuration (as generated by process \ref{proc:generate_binconfig}), and outputs correlated visibilities in the Flexible Image Transport System (FITS) format. Figure \ref{fig:correlation} shows the DFD for the workflow.

\subsubsection{Get start MJD}
\label{proc:get_start_mjd}
Because the voltage dump does not happen perfectly simultaneously across all the antennas, the data will have slightly different start times on a per-antenna basis. This process parses the voltage headers to find the earliest start time so that it can be provided to process \ref{proc:do_correlation} to be used as a reference time. This ensures that all the correlations are using the same reference time.

This process is executed a single time, and its output passed to each instance of process \ref{proc:do_correlation}.

\subsubsection{Do correlation}
\label{proc:do_correlation}
This process uses the ``DiFX" software correlator \citep{deller_difx-2_2011} to produce visibility datasets from the saved voltage data. One DiFX instance is executed for each of the 42 card-FPGA pairs, taking in two voltage files per antenna (one for each polarisation).  This produces 8 $\times$ 1.185 MHz output subbands (which are spaced by 1 MHz, but are wider than 1 MHz due to the ASKAP over-sampling), each with 128 frequency channels. The instance for the lowest-frequency card-FPGA pair is executed first, and its output passed to the other 41 instances (which are executed in parallel) to avoid issues caused by the FRB dispersion, as described below.

Because the 1~MHz-wide ASKAP coarse channels are over-sampled by a factor of 32/27, there is redundant data that can be discarded after correlation, and it is convenient to do so prior to the assembly of the DiFX output into FITS files.  We achieve this with a python script {\em mergeOverSampledDiFX.py}; from each 1~MHZ coarse channel, it retains 108 of the 128 frequency channels, corresponding to the non-overlapped portion of the band, and assembles the retained channels into a contiguous block.  Since each card-FPGA pair provides two blocks of four adjacent ASKAP coarse channels, the result is two 4~MHz subbands, no longer oversampled, each with 432 frequency channels.
 
If a bin configuration file, as generated by process \ref{proc:generate_binconfig}, is provided for the FRB, \change{one or more datasets of visibilities from only a subset of the available frequency-time range are produced, according to the definition in the bin config --- either a single RFI dataset or seven gate bin datasets (from the orange regions or within the green region depicted in Figure \ref{fig:bins_dynspec} respectively)}.  Due to the short length of the voltage files, the correction of the dispersive delays can lead to a frequency-dependent population of the output bin datasets --- at higher frequencies, the time correction can exceed the difference between the voltage file start and the FRB time at the low end of the band.  In this case, some bins may not have any visibility data for some frequencies, which leads to issues when subsequently assembling FITS files. To counter this, dummy data is generated for any baselines, times, and frequencies for which no visibilities were produced. \change{This dummy data has zero weight, and therefore does not impact the outputs down the processing chain.} The lowest frequency card-FPGA pair, which is guaranteed to have data due to having the latest FRB arrival time, is used as a template to provide the dummy data.

\subsubsection{Convert DiFX to FITS}
This process collects the output of all instances of process \ref{proc:do_correlation} to combine visibilities and convert the data into the FITS format. This produces a single FITS visibility file containing the full 336 MHz bandwidth, assembled into a single subband. During this process, frequency averaging is undertaken to reduce the data volume by a factor of 27, leading to a final spectral resolution of 250 kHz.

\subsection{Flag RFI-affected data}
Parts of the visibility data are often found to be corrupted by RFI or other systematic effects. Identification and flagging of corrupted visibilities are essential for calibration and imaging. This process performs data-flagging in three steps as described below \change{and is run post-correlation in the \workflow{fluxcal} and \workflow{polcal} workflows, and in the \workflow{FRB} workflow on the field mode correlated data}.
\begin{enumerate}
    \item Frequency channels that are known to be always affected by persistent RFI (from satellites) are flagged for all baselines.
    \item Data from each baseline are independently inspected for RFI affected channels. Identification of corrupted channels is performed based on the average (median) power and noise (median-absolute-deviation) in each channel. A frequency channel is identified as corrupted if its average power (or noise) is an outlier of the distribution for all channels. \change{The outlier threshold is calculated based on the number of data points and assuming Gaussian statistics. A nominal threshold is set at the value beyond which the number of expected data points drops below 1 for the given number of total data points and Gaussian statistics of the data. A multiplicative tolerance factor is applied to this nominal threshold to keep the flagging process conservative, especially in the initial rounds of flagging.} Since presence of RFI in a significant number of channels may bias the statistics, flagging is performed with a high tolerance factor in the first round. This step is repeated several times, \change{lowering the tolerance factor in each subsequent round. The tolerance factor used in the final rounds of flagging is close to unity.} 
    \item The statistics (average power and noise) of all baselines are compared together. Baselines having average power or noise which are outliers of the distribution for all baselines are identified as RFI-affected baselines, and flagged. An antenna is completely flagged if all its baselines are identified as RFI-affected baselines. \change{The outlier threshold for flagging is set in a similar manner as described in the previous point.}
\end{enumerate}
This process is independently executed on visibilities corresponding to the flux calibrator, the polarization calibrator and the field correlation. A flag file containing affected antennas, frequency channels, and/or baselines to be excised (on the basis of the above steps) is also generated for diagnostic purposes. Each of the calibration and imaging processes may also be provided an optional user-defined flag file if more flagging than is done automatically is required.

\subsection{Find flux calibration solutions}
\label{proc:find_flux_soln}

After flagging, we derive frequency-dependent complex gain solutions from the flux calibrator visibilities using three AIPS tasks: {\sc FRING} (solves for delay, i.e., a phase slope linearly proportional to frequency, with the solution amplitude fixed at unity), {\sc CPASS} (solves for frequency-dependent complex gain as a polynomial with frequency, normalising the average amplitude solution for each antenna to unity), and {\sc CALIB} (solves for a single frequency-independent complex gain per antenna, effectively setting the flux density scale and correcting for antenna-to-antenna signal level variations). While it would in principle be possible to combine these three solutions into a single stage, this separation allows for an easier identification of outliers based on delay and/or average amplitude correction.

We also apply these solutions to the flux calibrator visibilities themselves and convert to a CASA  measurement set for diagnostic purposes. The solutions are finally passed to the \workflow{polcal} and \workflow{FRB} workflows for imaging.

\subsection{Image polarisation calibrator}
\label{proc:image_polcal}

After flagging, we apply the delay and bandpass calibration tables as derived by process \ref{proc:find_flux_soln} to the polarisation calibrator visibilities using AIPS. We then convert the calibrated visibilities to a CASA measurement set and create an image of a 128" square region centred on the expected position of the polarisation calibrator with a 1" resolution using the CASA routine {\sc tclean}. \change{ASKAP's maximum angular resolution (with its longest baseline and at its highest frequency) is 6" \citep{hotan_australian_2021}, so our choice of 1" pixels is always sufficient.} We search the image for a single point source and fit its apparent position with the AIPS task {\sc JMFIT}, which is passed as an output to the \workflow{beamform} workflow.

\subsection{FRB localisation}

\subsubsection{Subtract RFI}
\label{proc:subtract_rfi}

Because FRB emission is often restricted to only a portion of the observing bandwidth and a significant amount of the signal is often in channels that are contaminated with RFI, we cannot simply flag channels as for the other visibility sets without losing significant signal-to-noise at best, or removing the FRB signal entirely at worst. Instead, we subtract the visibilities correlated in the RFI mode from those correlated in the gate mode, \change{without any time interpolation and} under the assumption that any RFI present is constant over the $\sim$50\,\si{\milli\second} surrounding the FRB. This process performs this subtraction, weighting the RFI visibilities by the ratio of the gate duration to the total duration of the RFI bins.

\change{We find that the noise post-RFI subtraction is consistent with white noise, and this method sufficiently removes RFI to make good images and localisations in most cases, but is occasionally imperfect. We treat imperfections by manually flagging data after RFI subtraction where necessary, and this is an area of ongoing improvement for CELEBI.}

\subsubsection{Image FRB}
\label{proc:image_FRB}
\begin{figure}
    \centering
    \includegraphics{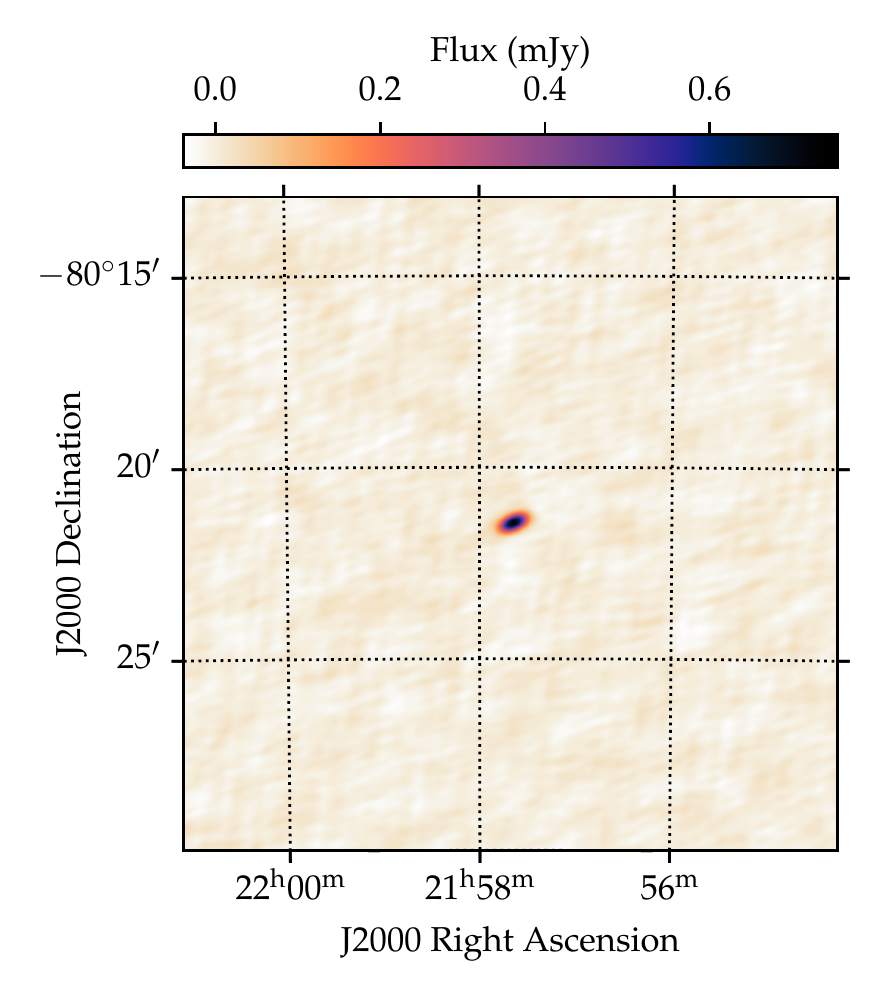}
    \caption{Image of FRB190711 created in process \ref{proc:image_FRB}.}
    \label{fig:190711_image}
\end{figure}

Using the RFI-subtracted visibilities from the FRB gate correlation, we calibrate, image, and fit the apparent position of the FRB in the same way as the polarisation calibrator (process \ref{proc:image_polcal}), but creating a 1024" square image due to the larger initial positional uncertainty. \change{As an example, an image of FRB190711 is shown in Figure \ref{fig:190711_image}.} The FRB's \change{measured} position \change{(right ascension $\mathrm{RA_{FRB}}$, declination $\mathrm{Dec_{FRB}}$)} is passed both directly to \workflow{beamform} and \change{with its error (right ascension $\Delta\mathrm{RA_{FRB}}$, declination $\Delta\mathrm{Dec_{FRB}}$)} to process \ref{proc:apply_offset}.

\subsubsection{Image field and find sources}
\begin{figure}
    \centering
    \includegraphics{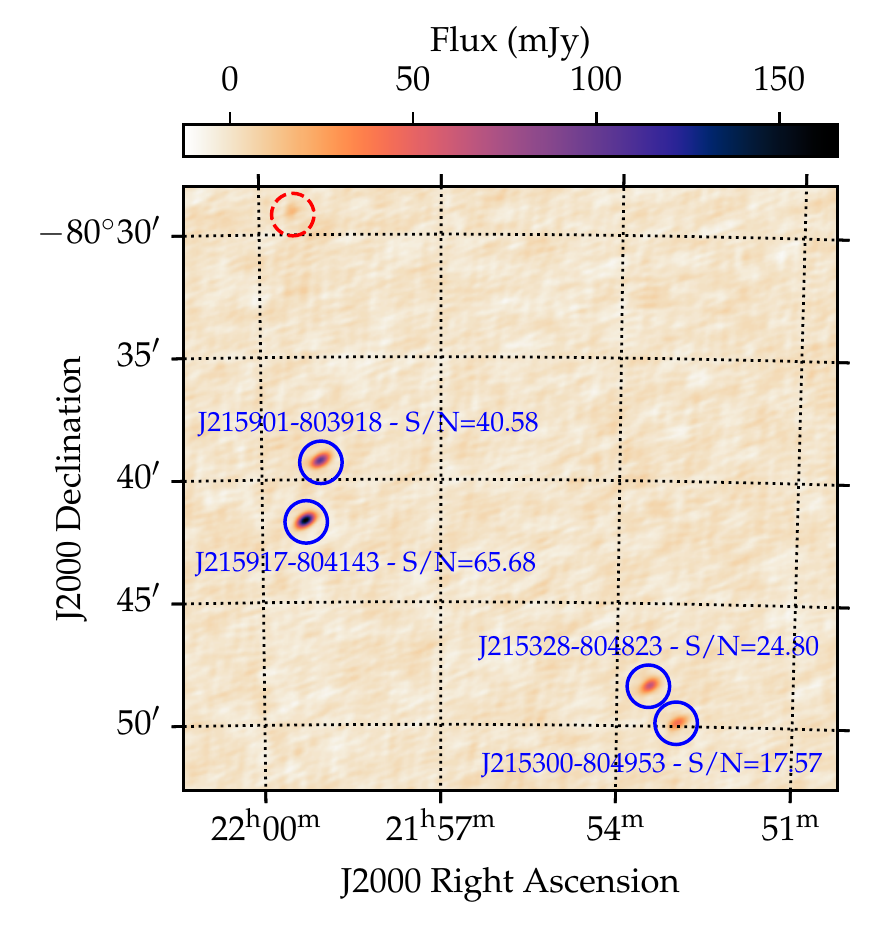}
    \caption{Cropped field image for FRB190711 created in process \ref{proc:image_field}. Blue circles indicate sources identified in the field image that returned one source in the RACS catalog, labeled with RACS component ID and detection S/N. The red dashed circle indicates an identified point source that did not pass the S/N threshold of 7 to be included in the offset analysis.}
    \label{fig:190711_field_image}
\end{figure}

\label{proc:image_field}
The field visibilities are calibrated, and imaged as for the polarisation calibrator and FRB, creating a 3000" square image. We identify up to 50 point sources in this image through the CASA task {\sc findsources} and fit their apparent positions with {\sc JMFIT}. These positions are then passed to process \ref{proc:find_offset}. \change{A cropped section of the field image created for FRB190711 with identified point sources marked is shown in Figure \ref{fig:190711_field_image}.}

\subsubsection{Find offset}
\begin{figure}
    \centering
    \includegraphics{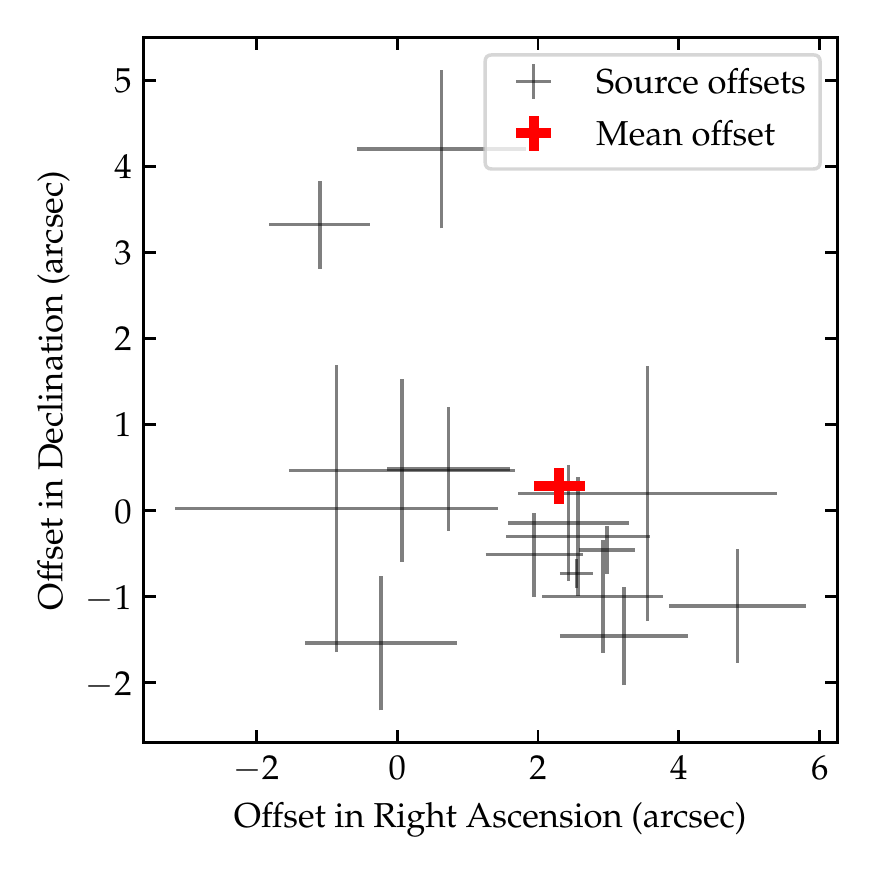}
    \caption{Offsets between measured field source positions and RACS catalog positions (thin grey) and mean offset as calculated by process \ref{proc:find_offset} (thick red) for FRB190711.}
    \label{fig:offsets}
\end{figure}
\label{proc:find_offset}
For each of the point sources identified in the field image, we \change{filter out those with measured S/N less than 7 and} search the Rapid ASKAP Continuum Survey (RACS) catalog \citep{hale_rapid_2021} with Astroquery for point sources within a 5" radius of the apparent position. \change{If more than one item is found in the catalog (which occurs with $\sim$1.4\% of our field sources), or if no items are found (which occurs with $\sim$22\% of our field sources) we discard the source. Identified sources discarded due to having no corresponding items returned from RACS are typically a result of {\sc findsources} incorrectly identifying a noise fluctuation as a source. In a small number of cases, sources in our image that appear to be real will not return any items from RACS, likely due to differences in observing frequency between the RACS catalog (which is in the low band) and our observation.} 

For each source in the remaining ensemble, we calculate the offset between our measured position and the RACS catalogue position, and then estimate a systematic positional correction (right ascension $\mathrm{\overline{RA}_{offset}}$, declination $\mathrm{\overline{Dec}_{offset}}$) \change{and error (right ascension $\Delta\mathrm{\overline{RA}_{offset}}$, declination $\Delta\mathrm{\overline{Dec}_{offset}}$)} for the field (and the FRB itself) using a weighted mean of these offsets multiplied by an empirical scaling factor (which accounts for differences in the angular resolution and frequency of our observations compared to the reference catalogue); the process is described in detail by \cite{day_astrometric_2021}.

\change{Figure \ref{fig:offsets} shows the offsets between CELEBI-measured positions and RACS catalog positions for the sources identified in the FRB190711 field image, as well as the calculated mean offset.}

\subsubsection{Apply offset}
\label{proc:apply_offset}
Finally, the mean offset and its error are added to the FRB's ASKAP position to obtain the corrected position (right ascension $\mathrm{RA_{corrected}}$, declination $\mathrm{Dec_{corrected}}$) and error:
\change{
\begin{align}
    \mathrm{RA_{corrected}}
        &= \mathrm{RA_{FRB}} + \frac{\mathrm{\overline{RA}_{offset}}}{\cos\left(\mathrm{Dec_{FRB}}\right)}
            \pm \sqrt{\Delta\mathrm{RA}_{\mathrm{FRB}}^2 + \Delta\mathrm{\overline{RA}}_{\mathrm{offset}}^2},\\
    \mathrm{Dec_{corrected}}
        &= \mathrm{Dec_{FRB}} + \mathrm{\overline{Dec}_{offset}}
            \pm \sqrt{\Delta\mathrm{Dec}_{\mathrm{FRB}}^2 + \Delta\mathrm{\overline{Dec}}_{\mathrm{offset}}^2}.
\end{align}
}

\begin{table}
\change{
\caption{Position-related quantities for FRB190711 as calculated by CELEBI.}
\label{tab:190711}
\centering
\begin{tabular}{ll} \hline
Quantity                                       & CELEBI-calculated value  \\\hline
$\mathrm{RA_{FRB}}$ (J2000, hh:mm:ss.s)        & 21:57:40.1 $\pm 0.2$     \\
$\mathrm{Dec_{FRB}}$ (J2000, dd:mm:ss.s)       & -80:21:29.1 $\pm 0.1$    \\
$\mathrm{\overline{RA}_{offset}}$ (arcsec.)    & $2.3\pm 0.3$             \\
$\mathrm{\overline{Dec}_{offset}}$ (arcsec.)   & $-0.4\pm0.2$             \\
$\mathrm{RA_{corrected}}$ (J2000, hh:mm:ss.s)  & 21:57:41.0 $\pm 0.4$     \\
$\mathrm{Dec_{corrected}}$ (J2000, dd:mm:ss.s) & -80:21:29.4 $\pm 0.3$    \\\hline
\end{tabular}
}
\end{table}

\change{Table \ref{tab:190711} lists the measured position, offset, and corrected position as calculated by CELEBI for FRB190711. We note agreement of the corrected position here with the position given by \cite{day_high_2020}.}
\section{Obtaining high-time resolution data via beamforming}
\label{sec:htr}

VCRAFT voltages can be used to reconstruct complex-valued time series of the electric field in the X and Y polarisations at the bandwidth-limited sample rate of $(336\,\si{\mega\hertz})^{-1}\approx3\,\si{\nano\second}$, coherently summed across antennas and coherently dedispersed to eliminate dispersion and associated smearing. This allows for construction of the Stokes parameters I, Q, U, V and measurements of the polarisation properties at high-time resolution and high S/N of FRBs detected and localised by ASKAP. Because we have access to the electric fields in X and Y directly, we can also construct arbitrarily-shaped dynamic spectra in I, Q, U, and V with freely-chosen time and frequency resolutions $\Delta t$ and $\Delta \f$, constrained only by $\Delta t \Delta \f \geq 1$. These dynamic spectra allow for polarimetric measurements across frequency and time, including the rotation measure (RM) and polarisation fractions.

In order to obtain these data products, the following operations must be performed on the voltages:
\begin{enumerate}
    \item Beamforming: the application of per-antenna time delays to account for the difference in signal arrival times due to the geometry of antennas and hardware signal propagation delays (process \ref{proc:do_beamform})
    \item PFB inversion: undoing the coarse channelisation performed by hardware before the voltages are recorded to obtain a single complex fine spectrum per polarisation per antenna (process \ref{proc:do_beamform})
    \item Calibration: the application of per-antenna bandpass calibration solutions, obtained during burst localisation, to the fine spectra (process \ref{proc:do_beamform})
    \item Summation: coherent summation of fine spectra across antennas to obtain a single fine spectrum per polarisation (process \ref{proc:sum})
    \item Derippling: removing systematic rippling in the fine spectra (processes \ref{proc:generate_deripple} and \ref{proc:apply_deripple})
    \item Coherent dedispersion (process \ref{proc:dedisperse})
    \item Inverse Fourier transform: obtain complex-valued time series at $(336\,\si{\mega\hertz})^{-1}\approx3\,\si{\nano\second}$, in the X and Y linear polarisation bases, via inverse Fourier transform of the fine spectra (process \ref{proc:ifft})
    \item Construct Stokes parameters and dynamic spectra (process \ref{proc:dynspecs})
\end{enumerate}

\subsection{Beamform}
\label{wf:beamform}

\begin{figure*}
    \centering
    \includegraphics[scale=0.5]{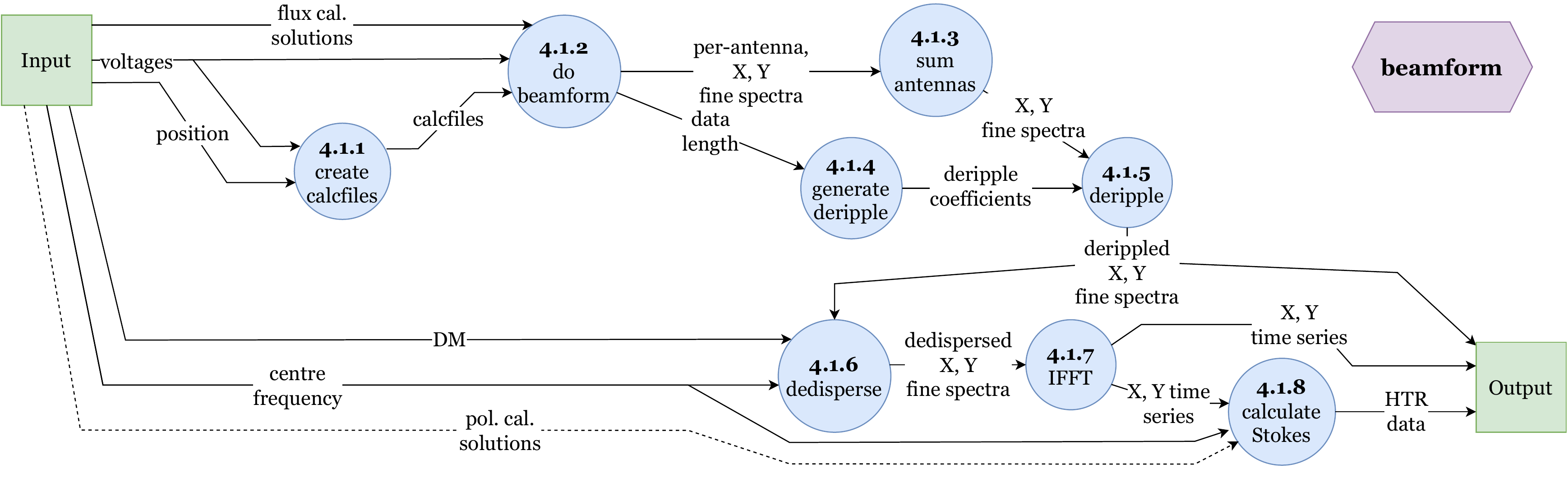}
    \caption{Beamform workflow DFD.}
    \label{fig:beamform}
\end{figure*}

The \workflow{beamform} workflow (Figure \ref{fig:beamform}) takes in a set of voltages, localised source position, flux calibration solutions, and optionally polarisation calibration solutions, and performs the operations listed above to produce a HTR data set, which includes: complex $\sim$3\,$\si{\nano\second}$-resolution time series in X and Y; Stokes I, Q, U, and V time series at the same time resolution; and arbitrarily-shaped I, Q, U, and V dynamic spectra (typically with $\Delta t=1\,\si{\micro\second}$ and $\Delta \f=1\,\si{\mega\hertz}$).

The \workflow{beamform} workflow is invoked within the \workflow{polcal} and \workflow{FRB} workflows, in both cases after the respective source has been localised. The position provided to the \workflow{FRB} instance of \workflow{beamform} is the apparent position, i.e. the position that is fit from the FRB image without astrometric correction.

The method for PFB inversion has also been described by \cite{morrison_performance_2020}, and the full method for obtaining high-time resolution FRB data by \cite{cho_spectropolarimetric_2020}. We describe these methods again here to reflect changes to the methods and describe the specific implementations applied in CELEBI.

\subsubsection{Create calcfiles}
\label{proc:create_calcfiles}
This process uses \texttt{difxcalc} to calculate the antenna-dependent geometric delays used in used by process \ref{proc:do_beamform} to align each antenna's datastream in time, given the previously-determined FRB position.

\subsubsection{Do beamform}
\label{proc:do_beamform}

This process prepares a fine spectrum (a spectrum of fine channels across the entire observing bandwidth) from voltages in each antenna for beamforming. This involves the application of antenna-dependent time delays, PFB inversion, and application of flux calibration solutions. An instance of this process is run for each unique polarisation-antenna pair, which we index with $p\in\{\mathrm{X}, \mathrm{Y}\}$ and $a \in [1\,..\,n_\mathrm{ant}]$ respectively, where $n_\mathrm{ant}$ is the number of antennas available.

The initial duration of the data across all antennas is equal, but because we are applying offsets to each antenna's data we must slightly reduce the duration of data loaded so that the data for all antennas occupies the same time range after applying the delays. We choose a number of samples $n_\mathrm{samp}$ that maximises the duration loaded while satisfying this condition.

In order to coherently sum signals across antennas, we must apply antenna-dependent time delays $\Delta t_a$ to the data. These have two components: a time-dependent geometric delay $G_a(t)$ that accounts for the differing path lengths of a signal between antennas; and a fixed delay $F_a$, which is measured during normal ASKAP operations and accounts for delays in hardware (mostly due to signal propagation delay in cables of different lengths for each antenna). \change{$F_a$ is measured every few days and is only expected to change when ASKAP's digitisers are reset, so we assume it to be stable between these resets.} These delays are provided by process \ref{proc:create_calcfiles}.

The geometric delay changes with time to account for the Earth's rotation changing the difference in arrival times over the duration of the 3.1-second voltage dump. \texttt{difxcalc} fits a polynomial to model the required geometric delay in each antenna as a function of time. Due to the short ($\sim$\,seconds) duration of data included in processing FRB voltages, the delay applied is well-approximated as being linear in time. $G_a(t)$ is evaluated via the interferometer model's polynomial at $t_\mathrm{start}$ and $t_\mathrm{end}$, corresponding to the start and end times of the data, and is linearly interpolated to give $G_a(t) \propto t$.

The coarse channelised voltages are loaded from all available VCRAFT files to obtain a complex time series for each coarse channel index $c \in [1\,..\,n_\mathrm{chan}]$ where in general the number of channels $n_\mathrm{chan} = 336$. When reading the data from disk, we load $n_\mathrm{samp}$ samples, offset from the beginning of the data by a number of samples equivalent to $F_a$. We then Fourier transform the complex time series to obtain a fine spectrum $s_{p;a;c}(\f)$. Each of these spectra are of oversampled coarse channels with central frequency $\f_c$. The geometric delay $G_a(t)$ is applied to $s_{p;a;c}(\f)$ to give an aligned spectrum:

\begin{equation}
    s^{\mathrm{align}}_{p;a;c}(\f) 
        = s_{p;a;c}(\f) e^{2i\pi \f_c G_a(t)}.
\end{equation}

By truncating each channel to remove the tapered regions and concatenating the flat regions of the fine spectra of the channels, we obtain a fine spectrum across the full bandwidth with a constant frequency response. First the truncation:

\begin{equation}
    s^{\mathrm{trunc}}_{p;a;c}(\f) 
        = 
    \begin{cases}
        0,               &   \f < \f_c - \frac{B_C}{2} \\
        s^{\mathrm{align}}_{p;a:c}(\f), &   \f_c - \frac{B_C}{2} \leq \f < \f_c + \frac{B_C}{2} \\
        0,               &   \f_c + \frac{B_C}{2} \leq \f
    \end{cases},
\end{equation}

\noindent and then the concatenation:

\begin{equation}
    S_{p;a}(\f) 
        = \sum_{c=1}^{n_\mathrm{chan}} s^{\mathrm{trunc}}_{p;a;c}(\f).
\end{equation}

\noindent Then we apply the flux calibration:

\begin{equation}
    S^\mathrm{cal}_{p;a}(\f)
        = P_{\mathrm{flux};a}(\f) S_{p;a}(\f),
\end{equation}

\noindent where $P_{\mathrm{flux};a}(\f)$ is an antenna-dependent phasor applying the flux calibration solutions as derived by process \ref{proc:find_flux_soln}.

\subsubsection{Sum antennas}
\label{proc:sum}

This process takes in the calibrated, beamformed fine spectra for each polarisation in each antenna output by process \ref{proc:do_beamform} and coherently sums these spectra to produce a single spectrum per polarisation:

\begin{equation}
    S_p (\f) 
        = \sum_{a=1}^{n_\mathrm{ant}} S^\mathrm{cal}_{p;a}(\f).
\end{equation}

\subsubsection{Generate deripple coefficients}
\label{proc:generate_deripple}

The design of ASKAP's PFB leads to the recovered fine spectra having a non-uniform, rippled frequency response (see Figure \ref{fig:oversamp_chan}). However, the exact shape of this rippling is predictable and it can be mitigated by dividing by a set of deripple coefficients. The deripple coefficients are the inverse of the coarse channel bandpass. This is determined by the FFT of the 24,576 ASKAP PFB coefficients $C_\mathrm{PFB}(\delta\f)$, where $\delta\f$ is frequency relative to the centre of the coarse channel, themselves a sinc function which is smoothed at the edges to reduce artefacts from the finite size of the filter. Fluctuations in the response are within 0.2\,dB over the nominal 1\,MHz coarse channel bandwidth \citep{tuthill_compensating_2015}. These coefficients are constant within the ASKAP system, and identical for each coarse channel and antenna. They have been generated once, and are hard-coded within CELEBI.

The derippling coefficients for each channel are

\begin{equation}
    C_\mathrm{derip}(\delta\f) = \frac{1}{\left|\mathcal{F}(C_\mathrm{PFB}(\delta\f))\right|}.
\end{equation}

\noindent Because the exact number of samples in the fine spectra ($n_\mathrm{samp}$) differs between input datasets, we linearly interpolate the denominator of this fraction to match the number of samples in each channel's truncated fine spectrum, i.e. $\lfloor n_\mathrm{samp} (B_C/B_{OS}) \rfloor$.

\subsubsection{Apply deripple coefficients}
\label{proc:apply_deripple}

Because the deripple coefficients $C_\mathrm{derip}(\delta\f)$ are identical for each coarse channel, we apply them by iterating over the central frequencies $\f_c$ of each of the $n_\mathrm{chan}$ coarse channels:

\begin{equation}
    S^\mathrm{derip}_p(\f_c+\delta\f)
        = S_p(\f_c+\delta\f)C_\mathrm{derip}(\delta\f).
\end{equation}

\noindent This produces fine spectra $S^\mathrm{derip}_p(\f)$ with uniform frequency responses.

\subsubsection{Coherently dedisperse}
\label{proc:dedisperse}

Dispersion is a well-modelled process, and one that is straightforward to account for in FRB data. Having access to the complex spectra of the X and Y polarisations enables coherent dedispersion, instead of imperfect incoherent dedispersion.

Coherent dedispersion is able to perfectly compensate for and remove the frequency-dependent time delay introduced by the ionised interstellar medium (assuming cold plasma dispersion) by acting on the voltage data that samples the electromagnetic wave in each of the two linear polarisations. This is because dispersion, as a physical process, effectively acts as a frequency-dependent rotation of phase in the spectral domain that manifests as a frequency-dependent time delay in the temporal domain. Therefore, with access to the spectral domain of the radiation being dispersed (the FRB signal), the phases can be de-rotated to obtain the signal as it would have been without any dispersion.  %(After signal detection in the real-time ICS search, the phase information is lost, and only incoherent dedispersion can be performed, at the frequency resolution available in the detected data products).

Assuming cold plasma dispersion, the transfer function for coherent dedipsersion to a dispersion measure DM is:

\begin{equation}
    H(\f;\,\mathrm{DM}) 
        = \exp\left(2i\pi k_\mathrm{DM}\mathrm{DM} \frac{(\f-\f_0)^2}{\f\f_0^2}\right),
\end{equation}

\noindent \citep{hankins_microsecond_1971} where $\f_0$ is a reference frequency, which we choose as the minimum frequency of the observing bandwidth. We apply this transfer function to the spectrum of each polarisation to coherently dedisperse them:

\begin{equation}
    S_{p;\mathrm{DM}}(\f) 
        = H(\f;\,\mathrm{DM}) S^\mathrm{derip}_p(\f).
\end{equation}

\subsubsection{Inverse fast Fourier transform}
\label{proc:ifft}

This process applies the inverse fast Fourier transform to the dedispersed fine spectra to obtain the complex electric field in each polarisation in the time domain:

\begin{equation}
    E_{p}(t) 
        = \mathcal{F}^{-1}\left(S_{p;\mathrm{DM}}(\f)\right).
\end{equation}

\subsubsection{Calculate Stokes parameters}
\label{proc:dynspecs}
We now calculate time series for the Stokes parameters:
\begin{align}
    I(t)
        &= |E_X(t)|^2 + |E_Y(t)|^2, \label{eq:stokes_I}\\
    Q(t)
        &= |E_X(t)|^2 - |E_Y(t)|^2, \\
    U(t)
        &= 2\operatorname{Re}(E_X^*(t)E_Y(t)),\\
    V(t)
        &= 2\operatorname{Im}(E_X^*(t)E_Y(t)). \label{eq:stokes_V}
\end{align}

\noindent The electric field time series can also be used to generate dynamic 
spectra with frequency resolution $\Delta\f$ and time resolution $\Delta t$ 
such that $\Delta\f\Delta t=1$. Typically, this is done with $\Delta\nu=1\,
\mathrm{MHz}\implies\Delta t=1\,\mathrm{\mu s}$, but is in general only 
constrained by $\Delta t = N_\mathrm{chan} \delta t$, where $N_\mathrm{chan}$ 
is a positive integer representing the number of channels desired in the dynamic 
spectra and $\delta t=(336\,\mathrm{MHz})^{-1} \approx 3\,\mathrm{ns}$ is the 
bandwidth-limited time resolution.

\begin{figure}
    \centering
    \includegraphics[width=\linewidth]{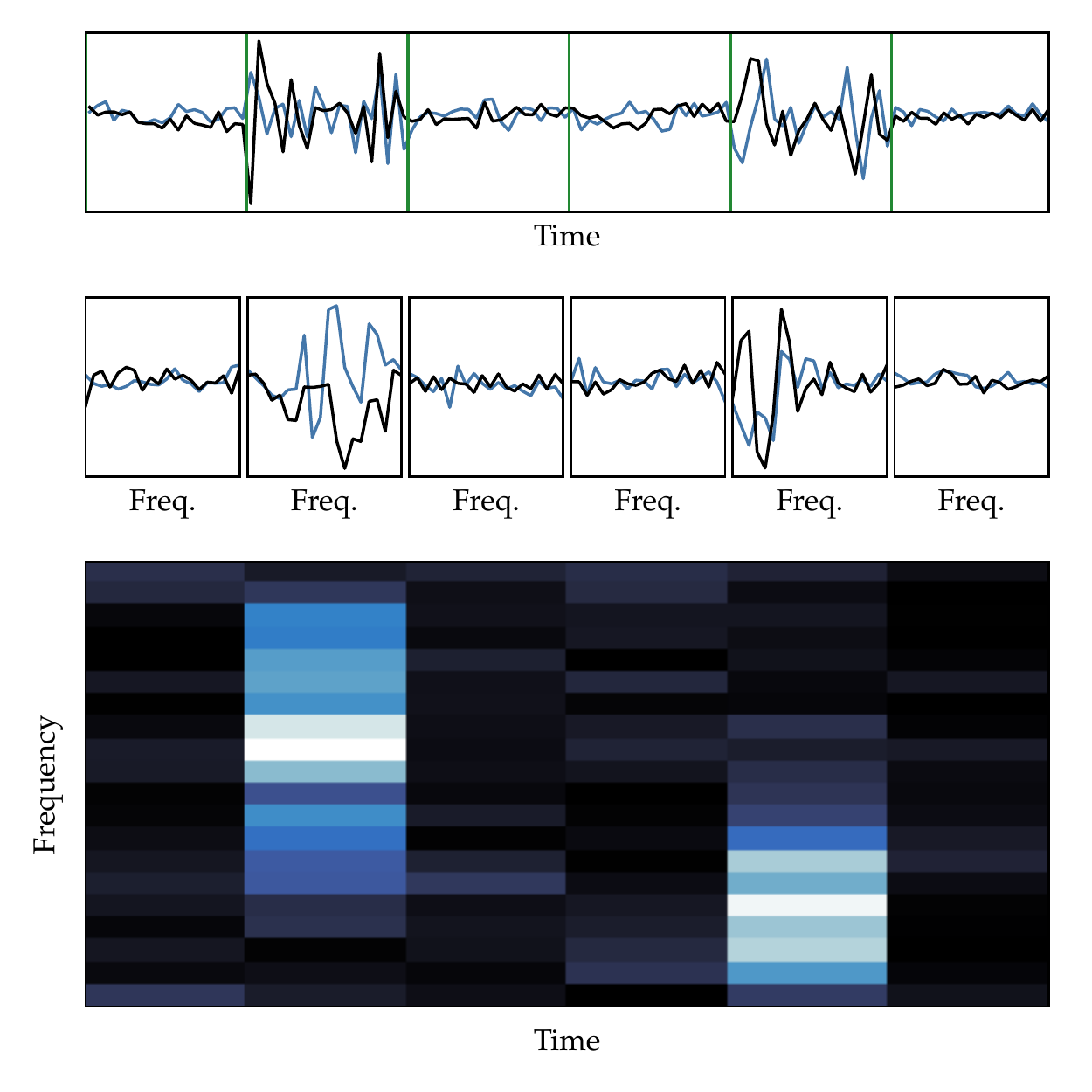}
    \caption{Visual representation of the process of converting a complex time
             series into a dynamic spectrum. Top panel: a simulated complex time series with the real component
             in black and the imaginary component in blue. Green lines separate
             sets of samples into bins of width $\Delta t$. 
             Middle panels: the complex Fourier transforms of each of the bins,
             again with the real component in black and imaginary component in 
             blue. Bottom panel: The amplitude of the dynamic spectrum created by plotting each 
             bin's spectrum vertically, with lighter cells representing higher 
             values.}
    \label{fig:dynamic_spectra}
\end{figure}

Once $\Delta\f$ and $\Delta t$ are selected, the dynamic spectra in each polarisation are generated by taking the discrete Fourier transform of
$N_\mathrm{chan}$ samples at a time. This process is demonstrated visually in 
Figure \ref{fig:dynamic_spectra}, and gives the dynamic spectra 
$E_X(t, \f)$ and $E_Y(t, \f)$. The Stokes dynamic spectra
are then calculated as before:

\change{
\begin{align}
    I(t, \f)
        &= |E_X(t, \f)|^2 + |E_Y(t, \f)|^2, \\
    Q(t, \f)
        &= |E_X(t, \f)|^2 - |E_Y(t, \f)|^2, \\
    U(t, \f)
        &= 2\operatorname{Re}(E_X^*(t, \f)E_Y(t, \f)),\\
    V(t, \f)
        &= 2\operatorname{Im}(E_X^*(t, \f)E_Y(t, \f)).     
\end{align}

We mitigate time-constant, frequency-dependent RFI in the dynamic spectra by zero-meaning and normalising each channel. For each channel in each Stokes dynamic spectrum, we zero-mean by subtracting the average value in that channel, and normalise by dividing by the standard deviation of the corresponding Stokes $I$ channel to ensure constant scaling between the Stokes parameters.
}

If polarisation calibration solutions as determined by process \ref{sec:polcal} have been provided, we also apply these to the Stokes U and V dynamic spectra:

\begin{align}
    U'(t,\f)
        &= U(t,\f)\cos(\Phi(\f)) - V(t,\f)\sin(\Phi(\f)), \\
    V'(t,\f)
        &= U(t,\f)\sin(\Phi(\f)) + V(t,\f)\cos(\Phi(\f)).
\end{align}

\subsection{Derive polarisation calibration solutions} 
\label{sec:polcal}

In order to correct for instrumental frequency-dependent leakage between Stokes $U$ and $V$, we take the Stokes dynamic spectra produced by process \ref{proc:dynspecs} for the polarisation calibrator data and derive a correction angle \change{$\Phi(\f)$} to apply to the FRB data using the method described by \cite{prochaska_low_2019}:

\begin{equation}
    \Phi(\f)
        = \Delta\tau\f + \Phi_0,
\end{equation}

\noindent where $\Delta\tau$ and $\Phi_0$ are leakage terms \change{respectively} representing a time and phase offset \change{between $U$ and $V$}. We derive models for the linear and circular polarisation ratios $L(\f)/I(\f)$ (where $L(\f) = \sqrt{Q(\f)^2 + U(\f)^2}$ \change{is the total linear polarisation}) and $V(\f)/I(\f)$ of each polarisation calibrator as second-order polynomials in $\f$ by fitting spectra obtained with the Murriyang radio telescope.

\subsection{Plot FRB high time resolution data}
\begin{figure*}
    \centering
    \includegraphics[width=\linewidth]{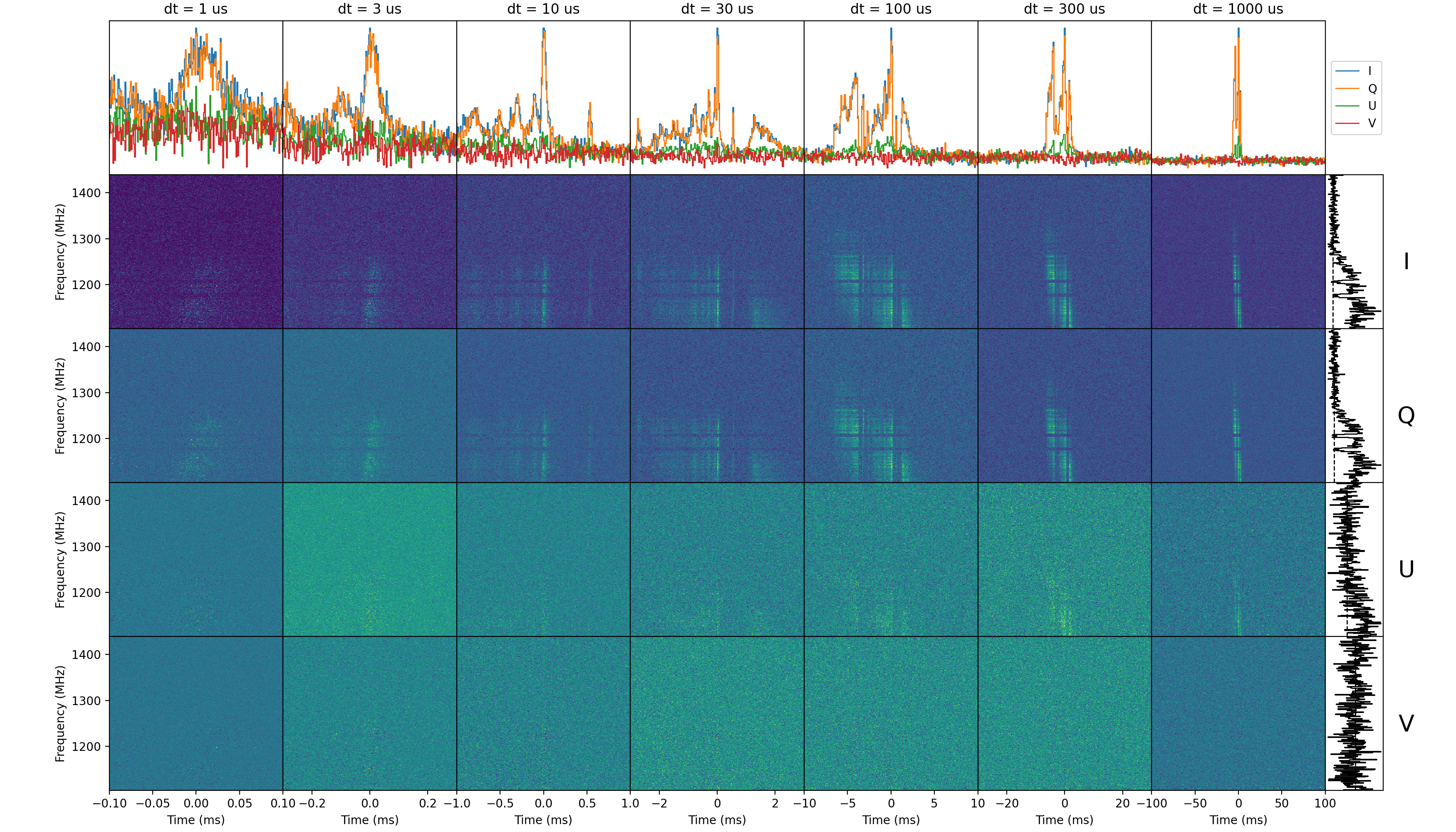}
    \caption{CELEBI output plot of Stokes dynamic spectra for FRB190711. Each row shows dynamic spectra for a Stokes parameter, labelled on the right, with the time averaging length labelled at the top of each column. The top row is the frequency-integrated pulse profile for each of the Stokes parameters. The rightmost column is the spectrum at the peak time index in the largest time averaging length.}
    \label{fig:210117_IQUV}
\end{figure*}

The final process of the high time resolution processing is to plot the data for the FRB. \change{Because the ideal time resolution for visual inspection of an FRB dynamic spectrum can be anywhere between 1\,\si{\micro\second} and 1\,\si{\milli\second}, and the sub-millisecond structure of the FRB is not known until this stage,} we plot each of the Stokes dynamic spectra over a range of time averaging values. \change{Figure \ref{fig:210117_IQUV} shows this plot as generated for FRB190711, and we note recovery of the high-time resolution structure reported by \cite{day_high_2020}.}

% \begin{figure*}
%     \centering
%     \includegraphics[width=\linewidth]{figures/210117_I_729.10.png}
%     \caption{CELEBI output plot of Stokes I time series for FRB210117, averaged to a time resolution of 1\,\si{\milli\second}.}
%     \label{fig:210117_I}
% \end{figure*}

\section{Summary}
\label{sec:discussion}

The bringing together of CRAFT's voltage processing software into CELEBI has led to several significant improvements to the software overall. Most importantly, voltage processing is now almost entirely automated. This has reduced the turnaround between FRB detection and obtaining the final data products (high-precision localisation and high-time resolution data) from a week or more of processing requiring a close level of human oversight and manual execution, to as little as less than a day with very little direct human supervision. The precise time required for processing depends on the resources available on the supercomputing cluster CELEBI is being run on (we have largely been using the OzStar supercomputer). Also, processing can be impeded by unexpected irregularities in the data or observations that CELEBI is not yet robust to. \change{Nextflow's handling of complex process relationships has greatly helped development, and its} method of process execution, where each instance of a process is executed in its own directory, combined with its detailed logging and reporting, makes diagnosis of problems quite straightforward and has greatly improved reproducability in processing. Processing CRAFT voltages is now much more accessible than it was previously, as the user-end interactions are much simpler and require less technical knowledge. \change{Responsive follow-up observations, on timescales on the order of a day, of CRAFT FRBs are now a possibility with the automation of procesing CRAFT voltages.}

A primary motivation for the development of CELEBI was the forthcoming CRACO upgrade for ASKAP's real-time detection system (Bannister et al., in prep). CRACO is expected to increase the rate at which ASKAP detects FRBs from of order $\sim$1 per month to of order $\sim$1 per day. This much higher detection rate will require automated processing and logging, and standardised data outputs, all of which are now provided via CELEBI.

While the primary functionality of CELEBI is now complete, development is ongoing. The processing of each FRB still requires a degree of human oversight, and processing errors are handled on a case-by-case basis. The robustness of CELEBI to issues such as data corruption, unusual antenna behaviour, calibration errors, and unexpected RFI environments is continually improving, but this can only occur as the issues arise. 

There are also remaining improvements to be made to optimise the quality of the pipeline outputs. The current method of imaging an FRB and subtracting RFI could be improved from the current binning (process \ref{proc:generate_binconfig}) by instead correlating of order 100 bins each $1\,\si{\milli\second}$ in duration, and using a matched filter post-correlation to image the burst while removing RFI. This would simplify the pipeline structure by making the RFI correlation branch obsolete, give more flexibility in imaging the FRB, and maximise the S/N of the FRB in its image, therefore minimising its positional uncertainty.

The high-time resolution data products are currently output as Numpy arrays. Incorporating conversion to other standard formats, such as PSRCHIVE archives or FITS, would be convenient for using the data products with already-existing analysis software. CELEBI also does not currently include any functionality for measuring FRB RMs, a burst property which is considered a standard measurement when possible, i.e. when polarimetric data is available.

\section*{Acknowledgements}

We thank Tyson Dial for comments on the manuscript. ATD and KG acknowledge support through ARC Discovery Project DP200102243. CWJ and MG acknowledge support from the Australian Government through the Australian Research Council's Discovery Projects funding scheme (project DP210102103). SB is supported by a Dutch Research Council (NWO) Veni Fellowship (VI.Veni.212.058). RMS acknowledges support through Australian Research Council Future Fellowship FT190100155 and Discovery Project DP220102305. This work was performed on the OzSTAR national facility at Swinburne University of Technology. The OzSTAR programme receives funding in part from the Astronomy National Collaborative Research Infrastructure Strategy (NCRIS) allocation provided by the Australian Government. This scientific work uses data obtained from Inyarrimanha Ilgari Bundara / the Murchison Radio-astronomy Observatory. We acknowledge the Wajarri Yamaji People as the Traditional Owners and native title holders of the Observatory site. CSIRO's ASKAP radio telescope is part of the Australia Telescope National Facility (\hyperlink{https://ror.org/05qajvd42}{https://ror.org/05qajvd42}). Operation of ASKAP is funded by the Australian Government with support from the National Collaborative Research Infrastructure Strategy. ASKAP uses the resources of the Pawsey Supercomputing Research Centre. Establishment of ASKAP, Inyarrimanha Ilgari Bundara, the CSIRO Murchison Radio-astronomy Observatory and the Pawsey Supercomputing Research Centre are initiatives of the Australian Government, with support from the Government of Western Australia and the Science and Industry Endowment Fund.

\bibliographystyle{elsarticle-harv} 
\bibliography{references}

\end{document}